%% file: combo.tex
\documentclass{mynature}

\usepackage{chapterbib}

\usepackage{graphics} 
\usepackage{times} 
\usepackage{mathptmx}

\usepackage[small,bf]{caption}

\typearea{12}

\renewcommand\refname{References}
\renewcommand{\thefootnote}{\fnsymbol{footnote}}

\usepackage{sectsty}

\begin{document}

\include{main}

\include{si}

\end{document}

%% file: main.tex

\bibliographystyle{naturemag}

\onecolumn

\section*{A low-frequency radio halo associated with a cluster of galaxies}

\noindent{\sffamily G. Brunetti$^1$,%
S. Giacintucci$^{1,2}$, %
R. Cassano$^{1}$,%
W. Lane$^{3}$, \newline
D. Dallacasa$^{4}$,%
T. Venturi$^{1}$, %
N. Kassim$^{3}$, %
G. Setti$^{1,4}$, \newline
W. D. Cotton$^{5}$, \& %
M. Markevitch$^{2}$}
\ \\

\noindent%
{\footnotesize\it%
$^{1}${INAF - Istituto di Radioastronomia, Via P. Gobetti 101,
  I-40129 Bologna, Italy}\\
$^{2}${Harvard-Smithsonian Center for Astrophysics, Cambridge, MA 02138, USA}\\
$^{3}${Naval Research Laboratory, Code 7213, Washington, DC 20375-5320, USA}\\
$^{4}${Dipartimento di Astronomia, Universita' di Bologna, Via Ranzani 1,
  I-40127 Bologna, Italy}\\
$^{5}${National Radio Astronomy Observatory, Charlottesville, Virginia
  22903-2475, USA}\\
}

\baselineskip26pt 
\setlength{\parskip}{12pt}
\setlength{\parindent}{22pt}%

\noindent{\bf 
Clusters of galaxies are the largest gravitationally bound objects in the
Universe, containing about $10^{15}$ solar masses of hot 
($10^8$ K) gas, galaxies
and dark matter in a typical volume of about 10 Mpc$^3$. 
Magnetic fields and relativistic particles are mixed with the gas as 
revealed by giant ‘radio haloes’, which arise from diffuse, megaparsec-scale 
synchrotron radiation at cluster center$^{1,2}$. 
Radio haloes require that the emitting electrons are
accelerated in situ (by turbulence)$^{3-6}$, or are injected (as secondary
particles) by proton collisions into the intergalactic 
medium$^{7-10}$. 
They are found only in a fraction of massive clusters that have complex
dynamics$^{11-14}$, which suggests a connection between these mechanisms and
cluster mergers. Here we report a radio halo at low frequencies associated
with the merging cluster Abell 521. This halo has an extremely steep radio
spectrum, which implies a high frequency cut-off; this makes the halo
difficult to detect with observations at 1.4 GHz (the frequency at which all
other known radio haloes have been best studied). The spectrum of the halo is
inconsistent with a secondary origin of the relativistic electrons, but
instead supports turbulent acceleration, which suggests that many radio
haloes in the Universe should emit mainly at low frequencies.}

The turbulent re-acceleration model$^{5, 6, 15-17}$ for the origin of 
giant radio haloes assumes that fossil relativistic particles are 
re-accelerated by
merger-induced turbulence to the energies necessary to produce the observed
radio synchrotron emission in relatively weak magnetic fields. 
The acceleration of fast particles by turbulence, known to be an important
process in astrophysics, is due to the resonant scattering of these
particles by the turbulent waves (the energy spectrum of the turbulence can
be thought of as the superposition of the contribution from turbulent waves
with different length scales), leading to a stochastic energization of
particles and to the damping of the waves$^{18}$. 
The connection between mergers
and particle acceleration by turbulence in galaxy clusters is complex. It is
argued$^{15, 19, 20}$ that it should take place on a
timescale of the order of a cluster–subcluster crossing time ($\sim$1 Gyr),
during which turbulence is continuously injected on scales of the order of
the subcluster size, transported at smaller scales and then dissipated into
heating of the intergalactic medium and acceleration of relativistic
particles over a fairly large volume. This argument is supported by recent
radio observations of a complete sample of X-ray luminous clusters that
allow a clear separation between clusters with radio haloes and radio-quiet
clusters, suggesting, from the fraction of clusters with radio haloes, that
the particle acceleration mechanisms operate sporadically, on timescales
$\leq$1 Gyr and in connection with cluster mergers$^{21}$.
Abell 521 is an X-ray luminous ($8.2 \times 10^{37}$ W 
in the 0.1–2.4-keV band) and
massive ($\sim 2 \times 10^{15}$ solar masses) galaxy cluster at 
redshift z = 0.247 with ongoing multiple merging 
episodes$^{22, 23}$. Here we report the discovery of a
giant radio halo in this cluster by means of deep observations with the
Giant Metrewave Radio Telescope (GMRT, India) at 240, 325 and 610 MHz. In
Fig. 1a we show the radio image of Abell 521 at 240 MHz, where the radio
halo is best imaged. To highlight the diffuse emission, point sources
visible in the full-resolution images were subtracted in producing the
low-resolution images in Fig. 1. The radio halo is coincident with the
cluster X-ray emitting region and correlates with the X-ray emission of the
hosting cluster, which is typical of other radio haloes$^{1,2}$ (see also
Supplementary Information).
Previous higher frequency Very Large Array (VLA) observations at 1,400 MHz
did not find this diffuse emission, instead observing only the radio relic
located on the southeastern boundary of the cluster$^{23}$. 
The relic coincides
with a possible shock front, generated by recent in-fall of a subcluster
along the northwest/southeast direction, where relativistic electrons are
currently accelerated$^{24}$. 
Figure 1 clearly shows that the radio halo becomes
progressively more dominant over the radio relic at lower frequencies,
indicating that its spectrum is much steeper than that of the relic, which
has $\alpha \approx 1.5$ (ref. 24; flux is proportional to $\nu^{-\alpha}$, 
where $\nu$ denotes frequency). 
The patchy structure of the radio halo at 610 MHz (Fig. 1b)
indicates the observational difficulty in imaging the emission, the surface
brightness of which is already fading at this frequency (note that
the largest diffuse emission detectable with the GMRT at 610 MHz 
is $\approx 0.28$ degrees, e.g. ref. 14). The halo disappears
between 610 and 1,400 MHz; only an upper limit on the flux of the radio halo
at 1,400 MHz can be derived, although faint residual emission in the cluster
is still present at this frequency (Fig. 1c and Supplementary Information).
The flux densities of the radio halo at 240, 325 and 610 MHz are plotted in
Fig. 2 together with the upper limits at 74 and 1,400 MHz. The upper limits
were evaluated by injecting fake radio haloes with different flux densities
into the observed data sets, following ref. 21, to estimate the sensitivity
of the observations to diffuse emission on the halo length scale. 
In particular, the upper limit at 74 MHz was derived from VLA Low-frequency 
Sky Survey data$^{25}$, whereas the upper limit at 1,400 MHz was derived 
from the analysis of the archival VLA data (Supplementary Information). 
The important
result is that the average value of the spectral index, $\alpha \approx 2.1$, 
is much
larger than that of any other known radio halo (typical spectral index is
$\alpha \approx 1.2$-$1.3$ (refs 1, 2)). 
These extreme spectral properties make Abell 521 a
unique system for addressing the origin of the emitting particles in radio
haloes.
Such an extremely steep spectrum and the downward spectral curvature 
(Fig. 2) imply a spectral cut-off at high frequency, 
which is a well-known signature of turbulent acceleration$^{4-6,16, 17, 19}$.
Synchrotron theory implies a corresponding cut-off in the spectrum of the
emitting electrons at $E_e \approx 1.4 B_{nT}^{-1/2}(\nu_c/300)^{1/2}$ GeV,
where $\nu_c$ is the cut-off frequency (measured in
megahertz) and $B_{nT}$ is the magnetic field (measured in nT). 
In the framework of the turbulent acceleration scenario, $E_e$
pinpoints the energy at which the timescale of the electron radiative losses
becomes equal to that of the turbulent acceleration. The timescale of
electrons emitting at $\nu_c$ is estimated taking into account the
redshift-dependent inverse-Compton losses against the cosmic microwave
background and the synchrotron losses$^{26}$, as follows, 
where $\nu_c$ is again
measured in megahertz:

\begin{equation}
\tau \approx 0.95
{{B_{nT}^{1/2} (\nu_c/300)^{-1/2} }\over
{(1+z)^4 + (B_{nT}/0.32)^2 }} \,\,\,\,\,\, {\rm Gyr}
\end{equation}

This means that the electrons responsible for the observed emission should
be accelerated on a timescale of $\sim (1.1-1.4) \times 10^8$ years, 
for 0.1 - 0.5-nT magnetic fields in the radio halo region. 
Assuming that fast magnetosonic
waves are responsible for the acceleration of the emitting particles,
following ref. 20 we find that this acceleration efficiency can be achieved
under the reasonable assumption that the energy density of these turbulent
waves is 12--18\% of the thermal energy (Fig. 2).

Appreciable synchrotron emission can also be produced by secondary electrons
injected by collisions between long-lived relativistic protons accumulated
in the cluster and the thermal protons in the intergalactic medium, and
secondary models have been proposed as alternatives to the re-acceleration
model to explain radio haloes$^{7 - 10}$. 
The very steep spectral slope of this
radio halo rules out secondary models by means of a straightforward energy
argument. To explain the spectrum of the radio halo through synchrotron
radiation from secondary electrons, the primary protons must have a very
steep spectral energy distribution $(N(p)\propto p^{-\delta}$, where   
$\delta \approx 4.2$ and $p$ denotes the particle momentum). 
The energy density of relativistic protons, $\epsilon_p$, required to 
match the observed synchrotron flux
through a secondary model can be estimated following the formalism 
in ref. 27. 
For an average number density of thermal protons of $n_{th}
\approx 1,500$ m$^{-3}$ in
the region of the radio halo (consistent with the average thermal density in
the same region derived from X-ray observations$^{22}$) and the 
synchrotron flux measured at 325 MHz, we find that $\epsilon_p$ ranges 
from approximately 3 to 100 times the energy density of the thermal
plasma for magnetic field values (averaged in the region of the radio halo)
ranging from B = 0.5 nT to B = 0.1 nT, respectively. 
This yields only a lower limit on the
energy density of high-energy protons, because for $\delta > 3$ 
an additional (even dominant) contribution to the energy comes from 
suprathermal particles with kinetic energies $<$1 GeV. 
A secondary origin for the emitting electrons thus
implies the unrealistic situation in which clusters are dominated by
non-thermal protons. It also violates present upper limits on the energy
density of these particles derived from $\gamma$-ray observations of several
clusters, at the 20\% level$^{28, 29}$.
If the spectrum of these protons has $\delta < 4$, then the synchrotron 
signal from
the secondary electrons produced by proton--proton collisions cannot exceed
the 1.4-GHz upper limit in Fig. 2, placing corresponding limits on the
energy density of the primary protons. Figure 3 shows upper limits on the
energy density of the primary protons in the region of the radio halo as a
function of the magnetic field. These limits were obtained following ref. 21
and show that the energy density of relativistic protons in the cluster is
less than 1\% of the thermal component for $B > 0.2$ nT and $\delta < 2.5$.
A similar conclusion has recently been reached in the analysis of a
statistical sample of galaxy clusters without radio haloes$^{21}$. 
However, in
that case a larger energy content of protons was still possible by assuming
that clusters without radio haloes have magnetic fields much smaller than
those with radio haloes. This alternative (ad hoc) possibility can be
reasonably ruled out in our case because Abell 521 hosts a radio halo with
bolometric radio luminosity comparable to that of classical radio haloes
($\nuP(\nu) \approx 10^{34}$ W). 
Future observations with the Fermi Gamma-ray Space
Telescope (formerly the Gamma-ray Large Area Space Telescope) 
will reveal galaxy clusters in $\gamma$-rays in cases where the energy content
of relativistic protons is significantly larger than about 1\% of the thermal
plasma. The combination of this future data and limits in the radio band
(Fig. 3 and ref. 21) may thus provide a powerful tool for constraining the
magnetic field strength in galaxy clusters.

As we look at applying the turbulent re-acceleration model to other galaxy
clusters, it should be stressed that the maximum energy to which electrons
can be re-accelerated and, ultimately, the cut-off frequency in the spectra
of radio haloes depend on the level of turbulence and on the properties of
the turbulent waves. The spectral cut-off affects our ability to detect
radio haloes in the Universe, introducing a strong bias against observing
them at frequencies substantially larger than $\nu_c$. 
Currently known radio
haloes are observed at gigahertz frequencies, requiring an efficient
turbulent acceleration mechanism. These haloes must result from the rare,
most energetic merging events and are therefore hosted only in the most
massive, hottest clusters$^{19,30}$, in line with 
observations$^{11 - 14}$. 
On the other hand, the majority of radio haloes should form during much more
common, but less energetic, merging events, for example between a massive
cluster and a substantially smaller subcluster (with mass ratio $>$5) or
between two similar clusters with mass $\leq 10^{15}$ solar masses$^{13,30}$. 
However,
these sources, with a cut-off in the synchrotron spectrum at $\nu_c < 1$ GHz,
should be visible only at lower frequencies, because their spectrum should
be similar to that of the low-frequency radio halo in Abell 521. Future
high-sensitivity radio telescopes operating at low frequencies, such as the
Low Frequency Array (LOFAR) and the Long Wavelength Array (LWA), are
expected to discover the majority of these sources and also to test their
connection with cluster mergers. At the moment this connection cannot be
tested, owing to the lack of observations of samples of galaxy clusters at
low radio frequencies.

\begin{figure}
\noindent\hspace*{-1.25cm}%
\resizebox{18.3cm}{!}{\includegraphics{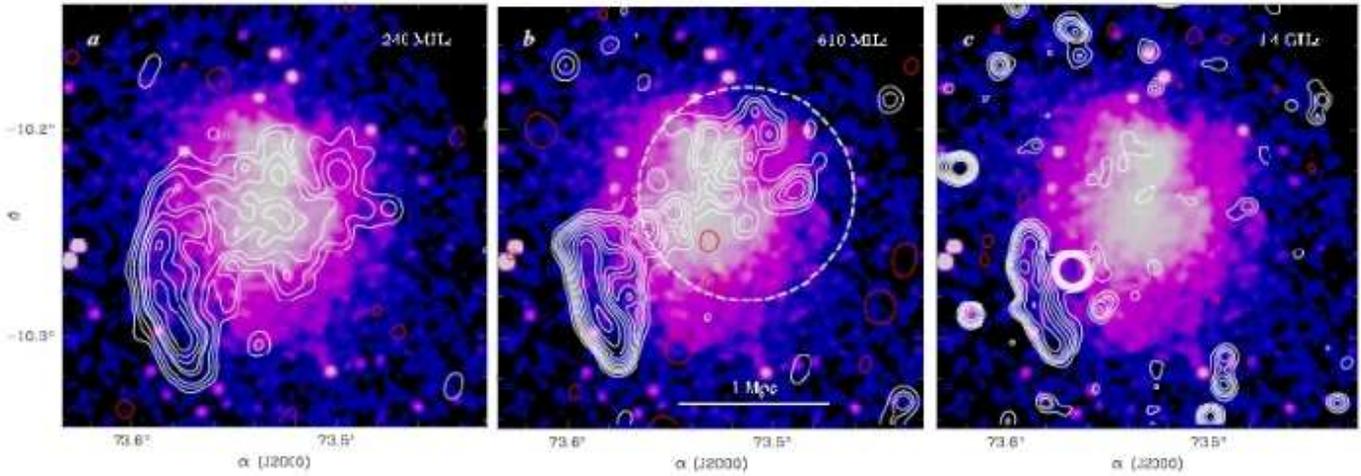}} %
\caption{
{\bf Radio and X-ray images of Abell 521.}
Low resolution radio contours (after subtraction of discrete
sources in the field identified at full resolution) overlaid 
on the Chandra 0.5-4 keV X--ray image (corrected for background and
exposure and smoothed with a $\sigma$=6 arcsec Gaussian, as in ref. 24).
{\bf Panel a)}: GMRT 240 MHz contours at
35 arcsec x 35 arcsec resolution.
The r.m.s. noise level in the cluster region is 220 $\mu$Jy/beam
(full resolution used for subtraction of discrete sources 
has 15.6 arcsec x 12.3 arcsec beam and r.m.s. noise =190 $\mu$Jy/beam).
Contours are at -0.66 (red), 0.66, 1.3, 2.6, 5.2, 7, 10, 14, 19 mJy/beam.
{\bf Panel b)}: GMRT 610 MHz contours at
35 arcsec x 35 arcsec resolution.
The r.m.s. noise level in the cluster region is 58 $\mu$Jy/beam
(full resolution used for subtraction of discrete sources 
has 9 arcsec x 4 arcsec beam and r.m.s. noise =35 $\mu$Jy/beam).
Contours are at -0.17 (red), 0.17, 0.34, 0.48, 0.67, 0.94, 1.3, 1.9, 2.6,
3.7, 5.2 mJy/beam.
The dashed circle indicates the region in which fluxes of the radio halo at 
different frequencies (reported in Fig.~2) are measured.
{\bf Panel c)}: VLA 1.4 GHz contours at 25 arcsec x 25 arcsec
resolution. 
Here only the discrete sources within 3 arcmin from the cluster centre 
are subtracted (other images are presented in Supplementary Information
together with a discussion on the subtraction of discrete sources at this
frequency).
The r.m.s. noise level in the cluster region is 26 $\mu$Jy/beam
(the full resolution used for subtraction of discrete sources
has 12.7 arcsec x 6.9 arcsec beam and r.m.s. noise =15 $\mu$Jy/beam).
Contours are at -0.08 (red), 0.08, 0.16, 0.22, 0.32, 0.45, 0.63, 0.89, 
1.25, 1.76, 2.48, 3.5, 4.9 mJy/beam. Coordinate system, J2000.
} 
\end{figure}

\begin{figure}
\begin{center}
\resizebox{12.0cm}{!}{\includegraphics{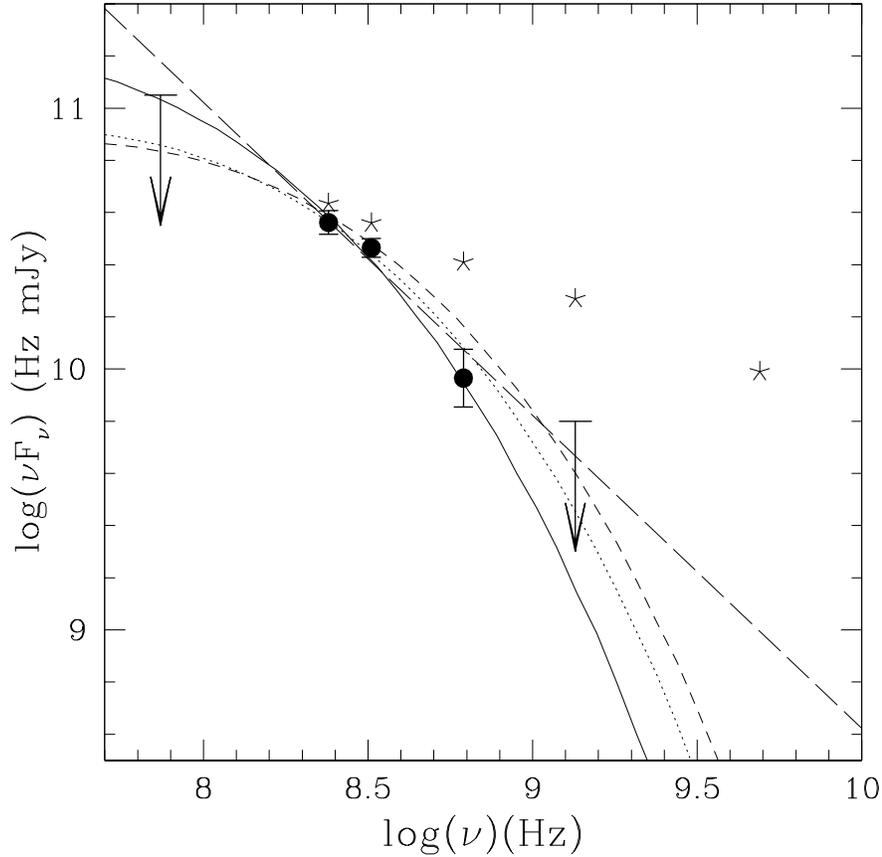}} %
\caption{
{\bf The spectrum of the radio halo.} Data are overlaid on secondary 
(long-dashed line) and re-acceleration models.
Measured fluxes ($F_{\nu}$) are : 152$\pm$15 mJy at 240 MHz,
90$\pm$7 mJy at 325 MHz, 15$\pm$3.5 mJy at 610 MHz, limits are 1.5 Jy at 74
MHz and 5 mJy at 1,400 MHz; errors (and error bars in the figure)
are given at 1 s.d.
For comparison, asterisks give the spectral energy distribution of
the radio relic (taken from ref.24).
The secondary model assumes $\delta =4.2$ and requires that
the energy density of relativistic protons is larger than that
of the thermal energy (see text).
Re-acceleration models assume : 14 \% of the thermal energy in
magnetosonic waves and a central value of the magnetic field
$B_o = 0.15$ nT (solid curve), 14 \% of the thermal energy in
magnetosonic waves and $B_o = $ 0.35 nT (dotted curve),
18 \% of the thermal energy in magnetosonic waves and
$B_o =$ 0.15 nT (short-dashed curve).
All the reacceleration models adopt a scaling 
$B \propto n_{{\rm th}}$ (e.g., ref.2 and refs therein).}
\end{center}
\end{figure}

\begin{figure}
\begin{center}
\resizebox{12.0cm}{!}{\includegraphics{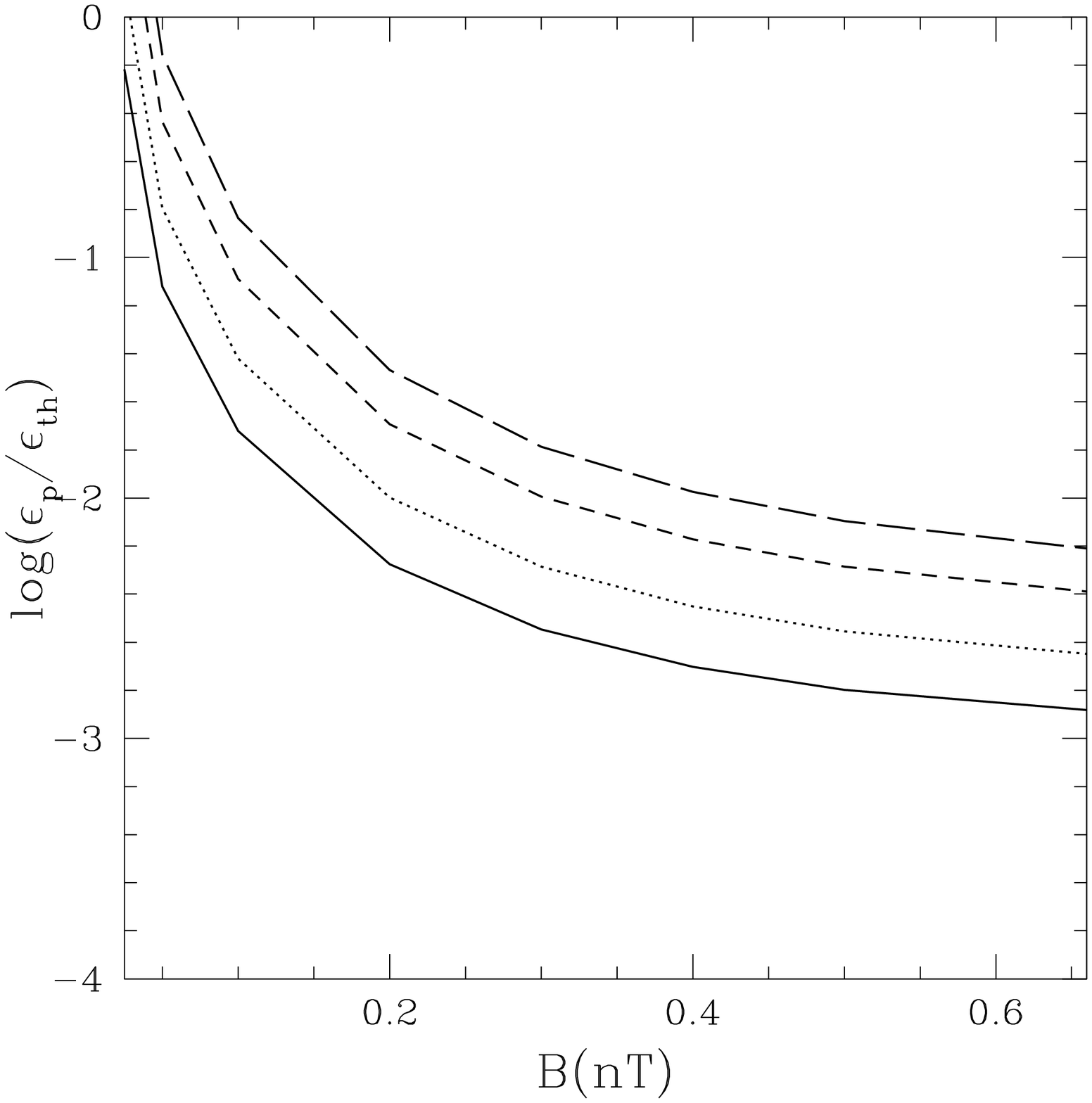}} %
\caption{
{\bf Limits to the energy density of relativistic protons.} 
Curves give the upper limit to the ratio between the energy density 
of relativistic protons and that of thermal gas, $\epsilon_{th}$, 
as a function of the 
magnetic field in Abell 521.
Calculations are shown assuming $n_{{\rm th}}=1,500$ m$^{-3}$
and $k_B T=$7 keV.
From bottom to top we adopt a spectra of the primary protons : 
$\delta$= 2.3, 2.5, 2.7, 2.9 ($N(p) \propto p^{-\delta}$).}
\end{center}
\end{figure}

\bibliography{main}

\vspace*{-0.5cm}\paragraph*{Acknowledgements} We acknowledge partial 
support from ASI-INAF I/088/06/0 and PRIN-INAF 2007.
G.B. and R.C. acknowledge the CfA, NRL and NRAO for hospitality and partial
support during the preparation of the manuscript.
Basic research in radio astronomy at the Naval Research Laboratory is
supported by 6.1 base funding. The National Radio Astronomy Observatory
is operated by Associated University, Inc., under cooperative
agreement with the National Science Foundation.
We thank R. Athreya 
the staff of the GMRT for their helping during the observations. 
GMRT is run by the National Center for Radio Tata Institute of Fundamental 
Research.

\vspace*{-0.5cm}\paragraph*{Author Information}
Reprints and permissions information is available at
npg.nature.com/reprintsandpermissions.                           
Correspondence and requests for materials should be addressed to
G. Brunetti (brunetti@ira.inaf.it).

\vspace*{-0.5cm}\paragraph*{Correspondence} and requests for materials should
be addressed to G.B.~(email: brunetti@ira.inaf.it).

%% file: si.tex

\paragraphfont{\small}
\subsectionfont{\normalsize}

\newcommand\be{\begin{equation}}
\newcommand\ee{\end{equation}}
\renewcommand{\vec}[1]{ {{\bf #1}} } 

\renewcommand{\textfraction}{0}
\renewcommand{\floatpagefraction}{1.0}
\renewcommand{\topfraction}{1.0}
\renewcommand{\bottomfraction}{1.0}

\renewcommand\caption[1]{%
  \myCaption{#1}
}

\title{\vspace*{-1cm}{\Large A low-frequency radio halo associated with a
cluster of galaxies}\vspace*{0.2cm} \\ {\em \large Supplementary Information}\vspace*{0.3cm}}

\author{\parbox{13.5cm}{\small\sffamily%
G. Brunetti$^1$,%
S. Giacintucci$^{1,2}$,%
R. Cassano$^{1}$,%
W. Lane$^{3}$, %
D. Dallacasa$^{4}$,%
T. Venturi$^{1}$, %
N. Kassim$^{3}$, %
G. Setti$^{1,4}$, %
W. D. Cotton$^{5}$,\& %
M. Markevitch$^{2}$}\vspace*{-0.5cm}}

\renewcommand\refname{\large References}

\date{}

\bibliographystyle{naturemag}

\baselineskip14pt 
\setlength{\parskip}{4pt}
\setlength{\parindent}{18pt}%

\twocolumn

\setlength{\footskip}{25pt}
\setlength{\textheight}{670pt}
\setlength{\oddsidemargin}{-8pt}
\setlength{\topmargin}{-41pt}
\setlength{\headsep}{18pt}
\setlength{\textwidth}{469pt}
\setlength{\marginparwidth}{42pt}
\setlength{\marginparpush}{5pt}

\addtolength{\topmargin}{-0.6cm}

\maketitle

\renewcommand{\thefootnote}{\arabic{footnote}}
\footnotetext[1]{\footnotesize INAF - Istituto di Radioastronomia, Via P. Gobetti 101, I-40129 
Bologna, Italy}
\footnotetext[2]{\footnotesize Harvard-Smithsonian Center for Astrophysics, Cambridge, 
MA 02138, USA}
\footnotetext[3]{\footnotesize Naval Research Laboratory, Code 7213, Washington, DC 20375-5320, USA}
\footnotetext[4]{\footnotesize Dipartimento di Astronomia, Universita' di Bologna, Via Ranzani 1,
I-40127 Bologna, Italy}
\footnotetext[5]{\footnotesize National Radio Astronomy Observatory, Charlottesville, Virginia
22903-2475, USA}

{\bf\small
To complement the information given in the main body of the Letter,
these Sections provide supplementary discussion on the 
methods adopted to derive upper limits to the flux density
of the halo reported 
in Figure 2 (Sect.~1) and on the origin of the halo (Sect.~2 and 3). \vspace*{-0.3cm}}

\renewcommand{\thefootnote}{\fnsymbol{footnote}}

\small

\subsection*{Upper limits to the flux of the radio halo at 1.4 GHz}

The diffuse emission in the center of Abell 521
discovered by the low frequency radio observations
presented in this Letter was not detected in VLA observations at 1.4
GHz$^{[17]}$.
We re-analyzed those BnC VLA data taken from the archive.
Given the sourthern declination of the cluster, they can
be considered equivalent to a C configuration observation. The quality
of the data is generally good, except for occasional intermittent
RFI in IF2 (1.44 GHz) and at the very end of the observation (both
25-MHz IFs affected), when the source and its secondary calibrator were
observed at low elevation and data on a few short baselines had
corrupted amplitudes (very high circular polarization).
RFI excision was carried out with great care.
We also paid particular attention to the analysis of the secondary
calibrator data, whose visibility turned out to be affected
by substantial contribution
from sources in the field. Therefore we first followed the standard
calibration procedure, then we used the image of the secondary
calibrator obtained in this way to improve the a-priori amplitude (and
phase) calibration accuracy. The final dataset was obtained after a
number of phase-only self-calibrations.
Solutions had generally very high signal to noise ratio.

The final images were obtained by considering a central square, 
0.8$^\circ$ wide, with additional 10 smaller (0.1$^\circ$) facets on
confusing sources.
In the full resolution image (12.7 arcsec $\times$6.9 arcsec in
p.a. 75$^\circ$) the r.m.s. noise level is 15 $\mu$Jy/beam.
We also produced images with lower resolution by applying natural
weighting to the data and also using a restricted uv-range.

A low resolution 30 arcsec$\times$30 arcsec image of the region of Abell 521
is shown in Figure S1.
Suspicious, very faint, diffuse emission appears in the region
of the radio halo discovered at lower frequencies; however, 
most of this emission is due to the convolution of point sources in
the same region (the image after subtraction of point sources is given 
in Figure 1) and the detection of the radio halo in Abell 521 at 1.4 GHz 
would not have been firmly established from these observations only.

\begin{figure*}
\begin{center}
\resizebox{\hsize}{!}{\includegraphics{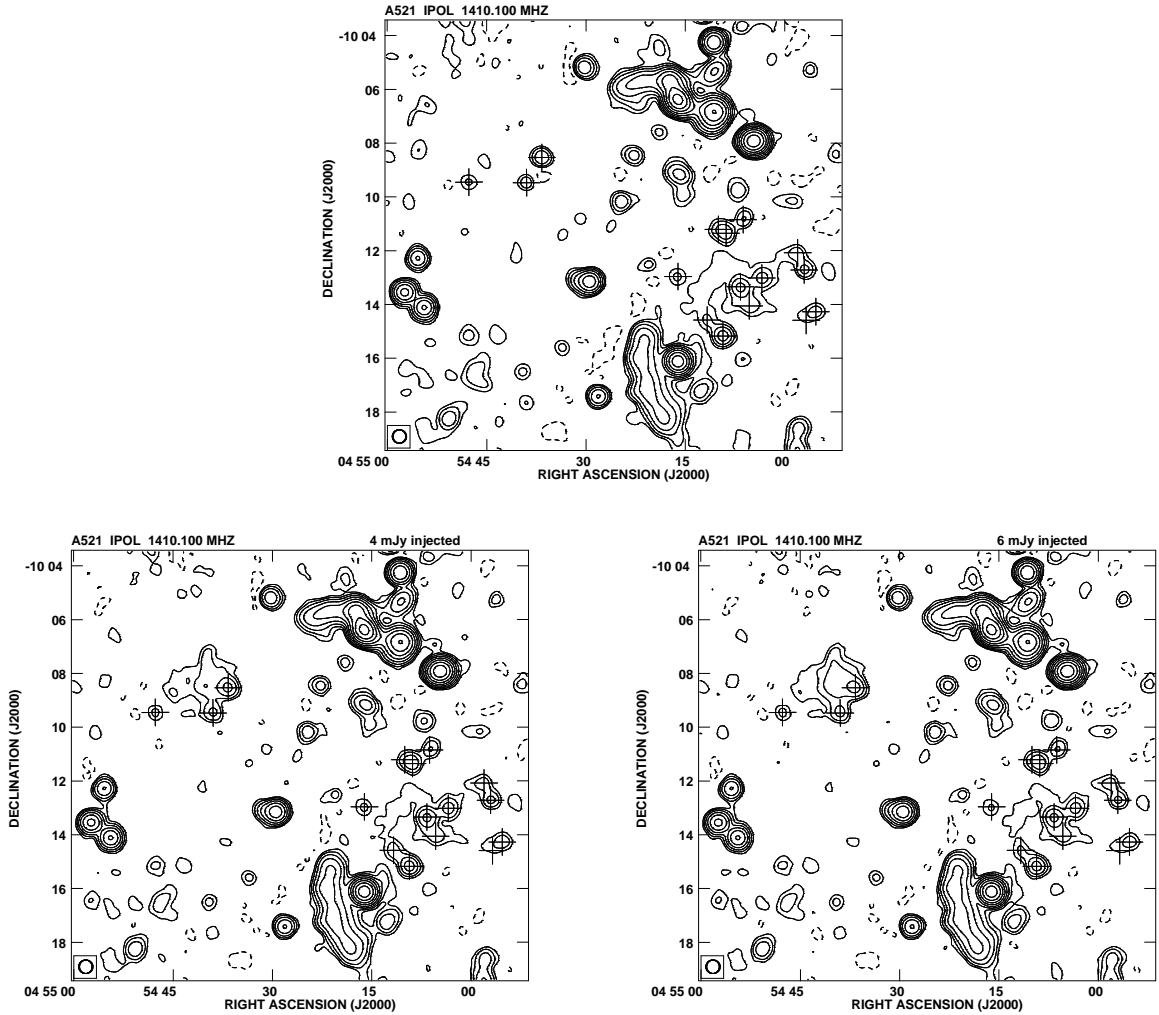}}
\end{center}
\caption[]{
{\bf Low resolution images of Abell 521 at 1.4 GHz.}
The upper panel shows the image from our analysis of the original data,
while lower panels show images from the analysis of the data containing
injected fake radio halos with $\Theta_{50\%} =$2.15 arcmin and $F$= 4 (left)
and 6 (right) mJy.
Crosses mark the position of the point sources identified in the high
resolution image in the central region of Abell 521 and in the region
where fake radio halos have been injected.
Radio contours are at 80, 160, 240, .. $\mu$Jy/beam level; 
negative contours at -80$\mu$Jy/beam level are reported in dashed lines.
The r.m.s. noise level in the images is in the range 28-30 $\mu$Jy/beam
and the restoring beam is 30 arcsec $\times$ 30 arcsec.}
\end{figure*}

We thus estimated the upper limit to the flux density of the radio halo
in these data at 1.4 GHz.
This should account also for the possibility that a fraction of the
flux density on the halo scale may be lost due to the
uv coverage at short baselines of the observations.
In order to derive a solid and conservative upper limit we
follow$^{[7,15]}$. We injected fake halos in
the uv--components of the dataset by means of the AIPS task UVMOD$^{[31]}$ 
and imaged the datasets
with the same set of parameters applied to the original data.
This procedure allows to evaluate the response of the 1.4 GHz VLA observation
to the presence of diffuse emission on the halo scale, as the uv
components of the fake radio emission are sampled by means of the real
uv coverage of the observation.

In particular,
we model the brightness profile of the fake halos with sets of optically thin
spheres with different radius and flux density, and obtain {\it families}
of fake radio halos with total flux densities ranging from 3 to 15 mJy,
with sizescale enclosing 50\% of the flux
$\Theta_{50\%} =$1.9 -- 2.3 arcmin ($\approx 440$ and $530$ kpc, respectively)
and maximum
angular extension $\approx 5.2$ arcmin ($\approx 1.2$ Mpc).

Figure S1 shows the results obtained for fake radio halos with
$\Theta_{50\%} =$2.15 arcmin and injected flux density $F$= 4 and 6 mJy.
In the case of the halo with $F$= 4 mJy,
about half of the injected flux is recovered
by the analysis of the dataset, and the resulting diffuse emission
in the image is still brighter than that at the center of Abell 521.

The analysis of the {\it families} of datasets with fake radio halos
allowed us to derive a conservative upper limit to the flux density at 
1.4 GHz of the radio halo in Abell 521, $F < 5$ mJy. 
This should be compared with the residual diffuse emission at
$F \approx 1-2$ mJy level measured in the region of the radio halo
after subtraction of discrete sources (Figure S1) in our images.
Because the flux density of these sources, $F=3-4$ mJy, is not negligible at 
1.4 GHz, they were fitted in the full resolution image (with beam = 
12.7 arcsec x 6.9 arcsec and r.m.s. =15 $\mu$Jy/beam) by using the AIPS 
task JMFIT solving for a Gaussian component (which has then been subtracted 
out) and a non-zero level (i.e. diffuse emission, which is not subtracted).

\subsection*{Projection effects and the radio halo}

Ghost radio plasma, confined within the intergalactic
medium, may be revived by the passage of a shock wave through adiabatic
compression$^{32}$; indeed this mechanism has been proposed
to explain radio relics as a scenario alternative to shock acceleration.

In principle, a large scale shock crossing Abell 521 along the line of sight
may compress ghost radio plasma producing detectable synchrotron
radiation with steep spectrum projected on the cluster central region, that
may be classified as radio halo.
However, the steep spectrum of the halo in Abell 521,
its large ($\approx$ Mpc) extension and - in particular - 
the morphological correlation
between radio halo and the X-ray emission from the cluster can be used
to conclude that this scenario is very unlikely.

In this context we have to assume that
the bubble of ghost plasma extends on the Mpc scale. 
This implies a dynamical age of this plasma
$\geq$ Gyr during which relativistic electrons lose energy via
radiative and adiabatic losses. This results in a cut-off in the synchrotron
spectrum emitted by the ghost plasma at very low frequencies, $\nu_{b,o}
\leq$MHz, before the shock--compression phase.
The passage of the shock
compresses the ghost plasma by a factor$^{32}$ $x = (P_{ds}/P_{us})^{3/4}$, 
where $P_{ds}$ and $P_{us}$ are the pressures measured
in the thermal medium around the ghost plasma downstream
and upstream the shock, respectively.
The adiabatic compression boosts the frequency of the cut-off
of the synchrotron spectrum at$^{33}$
$\nu_c \approx \nu_{c,o} x^{4/3} \approx P_{ds}/P_{us}$.
In order to have $\nu_c \approx 200-300$ MHz as measured
in the spectrum of the radio halo in Abell 521, this 
would require $P_{ds}/P_{us} > 300$ 
implying a Mach number of the shock (from the relation between pressure
jump and Mach number$^{34}$) :

$$M = {1\over{\sqrt{5}}}
\left(4 {{P_{ds}}\over{P_{us} }} +1 \right)^{1/2} > 15
\,\,\,\,\,\,\,\,\,\,\,\,\,\,\,\,\,\,\,\,\,\,\,\,\,\,\,\,\,\,\,\,\, (S.1)
$$

Strong large-scale shocks with $M > 10$ are rare in galaxy
clusters$^{35,36,37}$ and may occur only outside the virialised region
of the cluster where the temperature of the gas is well below 1 keV.
Thus the compression of the ghost plasma should happen at $\geq 3$ Mpc
distance from the center of Abell 521 (provided that a ghost-Gyr old
plasma at such distances manages to survive from mixing with 
the intergalactic medium).

At the same time the radio halo is correlated with the cluster
X-ray emission and its radio brightness decreases by a factor $\approx$6-8
on the scale of the core of Abell 521.
In order to reproduce this morphological connection by chance
the ghost radio
bubble, at a distance of $\geq 3$ Mpc from the cluster center, should
be aligned with the center of Abell 521 within a projected distance
of a cluster core-radius, $\approx 300$ kpc.
Geometrically, this implies that the direction connecting Abell 521
and the ghost bubble should make an angle $\theta 
\leq  7^o$ to the line of sight and the
chance probability for that, $\approx 1 - \cos(\theta)$, is $<$ 1\%.

The {\it ad hoc} assumptions and this low probability
allow us to conclude that this scenario is very unlikely.

\subsection*{On the size of the radio relic in Abell 521}

The radio halo in Abell 521 appears connected with the
radio relic (Figure 1).
This is similar to that observed in the case of
Abell 520$^{33}$ and Abell 2256$^{38}$; a bridge connecting the halo
and the relic is also observed in the Coma cluster$^{39}$.

Radio relics are interpreted as the result of the acceleration
of relativistic particles due to large scale shocks that
form in the intergalactic medium during cluster mergers
or accretion of matter$^{40,41}$.
The relic in Abell 521 is found to coincide with a possible
shock front with Mach number $M \approx 2-4$, generated by the recent infall
of a subcluster along the North-West/South-East direction$^{[18]}$.

Electrons accelerated at the shock flow away into the
intergalactic medium in a tail of emitting plasma,
and one might expect that part of the emission in the
region of the radio halo is due to this tail.
However in the following we show that this is unplausible : the extension
of the tail at a given frequency is constrained by the lifetime
of the electrons emitting at that frequency (that depends on inverse 
Compton and synchrotron losses, and on adiabatic losses),
which turns out to be much smaller than the $\geq$Mpc projected
distance between the relic and halo region.

We adopt the most favorable geometry assuming that the
shock is moving on the plane of the sky along the 
North-West/South-East direction, 
in which case the distance that separates the relic and halo region 
is not affected by projection effects.
We also assume that expansion in the shocked gas 
plays a role on a timescale longer than the radiative timescale of the 
electrons emitting at 240 MHz, that is $\approx 0.2$ Gyr (Eq.1), and
thus we neglect adiabatic losses of electrons.
This last assumption is justified since the gas in the
downstream region, over-pressured with respect to the ambient gas, 
should expand in a timescale that is a fraction of 
$\approx \phi_{r} / c_s \approx 0.5$ Gyr,
where $\phi_r \approx 500-700$ kpc is the longitudinal 
size of the relic$^{18}$ and $c_s$ the sound velocity.

Under these assumptions, 
the size of the tail at a given frequency
can be estimated as $l_{tail} \approx v_{ds} \tau_r$, where $\tau_r$
is the radiative
lifetime of electrons emitting at that frequency, and $v_{ds}$ is
the downstream velocity of the gas (in the shock frame, from the jump
conditions$^{34}$) :

$$
v_{ds} = c_s^u {{M^2 + 3}\over{4 M}}
\,\,\,\,\,\,\,\,\,\,\,\,\,\,\,\,\,\,\,\,\,\,\,\, (S.2)
$$

where $c_s^u$ is the sound velocity in the intergalactic medium
upstream of the shock. 
For a given Mach number of the shock $c_s^u$ 
can be estimated from the temperature of Abell 521
(taken as downstream temperature, and from the relations 
between temperature jump and Mach number$^{34}$) :

$$
c_s^u \approx 1700 \sqrt{ {{16 M^2}\over{(5 M^2 -1)(M^2+3)}} }
\,\,\,\,\,\, {\rm (km/s)} \,\,\,\,\,\,\,\,\,\,\,\,\,\,\,\,\,\, (S.3)
$$

Combining S.2--S.3 and adopting $\tau_r \approx 0.2$ Gyr for the electrons
emitting at 240 MHz (Eq.~1), we obtain
$l_{tail} \approx 200$ kpc if the relic in Abell 521 in generated by a shock
with $M \approx 2-4$.

\noindent
Remarkably, $l_{tail} \approx 200$ kpc agrees with the
transverse size of the
relic at 240 and 330 MHz, and with the size of the synchrotron spectral
steepening in the relic as measured from radio observations$^{18}$.

\bibliography{si}